\documentclass[twocolumn,preprintnumbers,prd,superscriptaddress,nofootinbib]{revtex4}

\usepackage{graphicx}
\usepackage{epsf}
\usepackage{bm}
\usepackage{amsmath}
\usepackage{amsfonts}
\usepackage{amssymb}
\usepackage{epstopdf}
\usepackage{natbib}
\usepackage{hyperref}
\usepackage{color}
\usepackage{verbatim}
\usepackage{multirow}
\usepackage{bm}
\usepackage{hyperref}

\begin{document}

\title{Measurements of $H_0$ and reconstruction of the dark energy properties from a model-independent joint analysis}

\author{Alexander Bonilla}
\email{abonilla@fisica.ufjf.br}
\affiliation{Departamento de F\'isica, Universidade Federal de Juiz de Fora, 36036-330, Juiz de Fora, MG, Brazil}

\author{Suresh Kumar}
\email{suresh.kumar@pilani.bits-pilani.ac.in}
\affiliation{Department of Mathematics, National Institute of Technology, Kurukshetra, Haryana-136119, India}
\affiliation{Department of Mathematics, BITS Pilani, Pilani Campus, Rajasthan-333031, India}

\author{Rafael C. Nunes}
\email{rafadcnunes@gmail.com}
\affiliation{Divis\~ao de Astrof\'isica, Instituto Nacional de Pesquisas Espaciais, Avenida dos Astronautas 1758, S\~ao Jos\'e dos Campos, 12227-010, SP, Brazil}

\begin{abstract}
Gaussian processes (GP) provide an elegant and model-independent method for extracting cosmological information from the observational data. In this work, we employ GP to perform a joint analysis by using the  geometrical cosmological probes such as Supernova Type Ia (SN), Cosmic chronometers (CC), Baryon Acoustic Oscillations (BAO), and the H0LiCOW lenses sample to constrain the Hubble constant $H_0$, and reconstruct some properties of dark energy (DE), viz., the equation of state parameter $w$, the sound speed of DE perturbations $c^2_s$, and the ratio of DE density evolution $X = \rho_{\rm de}/\rho_{\rm de,0}$. From the joint analysis SN+CC+BAO+H0LiCOW, we find that $H_0$ is constrained at 1.1\% precision with $H_0 = 73.78 \pm 0.84$ km s$^{-1}$Mpc$^{-1}$, which is in agreement with SH0ES and H0LiCOW estimates, but in $\sim$6.2$\sigma$ tension with the current CMB measurements of $H_0$. With regard to the DE parameters, we find $c^2_s < 0$ at $\sim$2$\sigma$ at high $z$, and the possibility of $X$ to become negative for $z > 1.5$. We compare our results with the ones obtained in the literature, and discuss the consequences of our main results on the DE theoretical framework.
\end{abstract}

%\keywords{}

%\pacs{}

\maketitle

\section{Introduction}

Several astronomical observations indicate that our Universe is currently in an accelerated expansion stage \cite{A_01,A_02,A_03,A_04,A_05}. A cosmological scenario with cold dark matter (CDM) and dark energy (DE) mimicked by a positive cosmological constant, the so-called $\Lambda$CDM model, is considered the standard cosmological model, which fits the observational data with great precision.
But, the cosmological constant suffers from some theoretical problems \cite{DE_01,DE_02,DE_03}, which motivate alternative considerations that can explain the data and have some theoretical appeal as well. In this regard, numerous cosmological models have been proposed in the literature, by introducing  some new dark fluid with negative pressure or modification in the general relativity theory, where additional gravitational degree(s) can generate the accelerated stage of the Universe at late times (See \cite{DE_review01,DE_review02,DE_review03} for a review). On the other hand, from an observational point of view, it is currently under discussion whether the $\Lambda$CDM model really is the best scenario to explain the observations, mainly in light of the current Hubble constant $H_0$ tension. Assuming the $\Lambda$CDM scenario, Planck-CMB data analysis provides $H_0=67.4 \pm 0.5$ km s$^{-1}$Mpc$^{-1}$ \cite{Planck2018}, which is in $4.4\sigma$ tension with a cosmological model-independent local measurement $H_0 = 74.03 \pm 1.42$ km s$^{-1}$Mpc$^{-1}$ \cite{R19} from the Hubble Space Telescope (HST) observations of 70 long-period Cepheids in the Large Magellanic Cloud. Additionally, a combination of time-delay cosmography from H0LiCOW lenses and the distance ladder measurements are in $5.2\sigma$ tension with the Planck-CMB constraints \cite{H0LiCOW} (see also \cite{H0LiCOW_2} for an update using H0LiCOW lens based  new hierarchical approach where the  mass-sheet transform is only constrained by stellar kinematics). Another accurate independent measure was carried out in \cite{Freedman}, from Tip of the Red Giant Branch, obtaining $H_0 = 69.8 \pm 1.1$ km s$^{-1}$Mpc$^{-1}$. Several other estimates of $H_0$ have been obtained in the recent literature (see \cite{Hubble_1,Hubble_2,Hubble_3,Hubble_4,Hubble_5}). It has been widely discussed in the literature whether a new physics beyond the standard cosmological model can solve the $H_0$ tension \cite{H0_1,H0_2,H0_3,H0_4,H0_5,H0_6,H0_7,H0_8,H0_9,H0_10,H0_11,H0_12,H0_13,H0_14,H0_15}. The so-called $S_8$ tension is also not less important. It is present between the Planck-CMB data with respect to weak lensing measurements and redshift surveys, about the value of the matter energy density $\Omega_m$ and the amplitude or growth rate of structures ($\sigma_8$, $f\sigma_8$). We refer the reader to \cite{S8_tension_1,S8_tension_2} and references therein for perspectives and discussions on $S_8$ tension.  Some other recent studies/developments \cite{D_1,D_2,D_3,D_4,D_5,D_6,D_7,D_8,D_9,D_10,D_11,D_12} also suggest that the minimal $\Lambda$CDM model is in crisis.

A promising approach for investigation of the cosmological parameters is to consider a model-independent analysis. In principle, this can be done via cosmographic approach \cite{cosmographic_01, cosmographic_02,cosmographic_03,cosmographic_04,cosmographic_05}, which consists of performing a series expansion of a cosmological observable around $z=0$, and then using the data to constrain the kinematic parameters. Such a procedure works well for lower values of $z$, but can be problematic at higher values of $z$. An interesting and robust alternative can be to consider a Gaussian process (GP) to reconstruct cosmological parameters in a model-independent way. The GP approach is a generic method of supervised learning (tasks to be learned and/or data training in GP terminology), which is implemented in regression problems and probabilistic classification. A GP is essentially a generalisation of the simple Gaussian distribution to the probability distributions of a function into the range of independent variables. In principle, this can be any stochastic process, however, it is much simpler in a Gaussian scenario and it is also more common, specifically for regression processes, which we use in this study. The GP also provides a model independent smoothing method that can further reconstruct derivatives from data. In this sense, the GP is a non-parametric strategy because it does not depend on a set of free parameters of the particular model to be constrained, although it depends on the choice of the covariance function, which will be explained in more detail in the next section. The GP method has been used to reconstruct the dynamics of the DE, modified gravity, cosmic curvature, estimates of Hubble constant, and other perspectives in cosmology by several authors \cite{GP_01,GP_02,GP_03,GP_04,GP_05,GP_06,GP_07,GP_08,GP_09,GP_10,GP_11,GP_12,GP_13,GP_14,GP_15,GP_17,GP_18,GP_19,GP_20,GP_21,GP_22,GP_23}.

In this work, our main aim is to employ GP to perform a joint analysis by using the  geometrical cosmological probes such as Supernova Type Ia (SN), Cosmic chronometers (CC), Baryon Acoustic Oscillations (BAO), and the H0LiCOW lenses sample to constrain the Hubble constant $H_0$, and reconstruct some properties of DE, viz., the equation of state parameter $w$, the sound speed of DE perturbations $c^2_s$, and the ratio of DE density evolution $X = \rho_{\rm de}/\rho_{\rm de,0}$. These are the main quantities that can represent the physical characteristics of DE, and possible deviations from the standard values $w=-1$, $c^2_s = 1$ and $X=1$, can be an indication of a new physics beyond the $\Lambda$CDM model. To our knowledge, a model-independent joint analysis from above-mentioned data sets, as will be presented here, is new and not previously investigated in the literature. Indeed, a joint analysis with several observational probes is helpful to obtain tight constraints on the cosmological parameters.

This paper is structured as follows. In Section \ref{Methodology}, we present the GP methodology as well as the data sets used in this work. In Section \ref{results}, we describe the modelling framework providing the cosmological information, and discuss our main results in detail. In Section \ref{conclusion}, we summarize main findings of this study with some future perspectives. \\

\section{Methodology and data analysis}
\label{Methodology}

In this section, we summarize our methodology as well as the data sets used for obtaining our results.

\subsection{Gaussian Processes}\label{GaP}

The main objective in a GP approximation is to reconstruct a function $f(x_i)$ from a set of its measured values $f(x_i) \pm \sigma_i$, where $x_i$ represent the training points or the positions of the observations. It assumes that the value of the function at any point $x_i$ follows a Gaussian distribution. The value of the function at $x_i$ is correlated with the value at other point $x_i'$. Therefore, we may write the GP as

\begin{equation}
\label{eqn:GPs}
f(x_i)=\mathcal{GP}(\mu(x_i),\textrm{cov}[f(x_i),f(x_i)]),
\end{equation}
where $\mu(x_i)$ and $\textrm{cov}[f(x_i),f(x_i)]$ are the mean and the variance of the random variable at $x_i$, respectively. This method has been used in many studies in the context of cosmology (e.g. see \cite{GP_01,GP_02,GP_03}). For the reconstruction of the function $f(x_i)$, the covariance between the values of this function at different positions $x_i$ can be modeled as
\begin{equation}
\label{eqn:cov}
\textrm{cov}[f(x),f(x')] = k(x,x'),
\end{equation}
where $k(x,x')$ is a priori assumed covariance model (or kernel in GP language), and its choice is often very crucial for obtaining good results regarding the reconstruction of the function $f(x_i)$. The covariance model, in general, depends on the distance  $|x-x'|$ between the input points ($x, x'$), and the covariance function $k(x,x')$ is expected to return large values when the input points ($x, x'$) are close to each other. The most popular and commonly used covariance functions in the literature are the standard Gaussian Squared-Exponential (SE) and the Mat\'ern class of kernels ($M_{\nu}$). The SE kernel is defined as

\begin{equation}
\label{eqn:kSE}
k_{SE}(x,x') = \sigma_f^2 \exp\left(-\frac{|x-x'|^2}{2 l^2}\right),
\end{equation}
where $\sigma_f$ is the signal variance, which controls the strength of the correlation of the function, and $l$ is the length scale that determines the ability to model the main characteristics (global and local) in the evaluation region to be predicted (or coherence length of the correlation in $x$). These two parameters are often called hyperparameters. They are not the parameters of the function, but of the covariance function. For convenience, in what follows, we redefine $\tau = |x-x'|$, which is consistent with all the kernels implemented here. The SE kernel, however, is a very smooth covariance function which can very well reproduce global but not local characteristics. To avoid this, the Mat\'ern class kernels are helpful, and the general functional form can be written as

\begin{equation}
    k_{M_{\nu}}(\tau) = \sigma_f^2 \frac{2^{1-\nu}}{\Gamma(\nu)} \left( \frac{\sqrt{2 \nu}\tau}{l} \right)^{\nu} K_{\nu}\left( \frac{\sqrt{2 \nu}\tau}{l} \right),
\end{equation}
where $K_{\nu}$ is the modified Bessel function of second kind, $\Gamma(\nu)$ is the standard Gamma function and $\nu$ is strictly a positive parameter. An explicit analytic functional form for half-integer values of $\{\nu = 1/2, 3/2, 5/2, 7/2, 9/2, .. \}$ is provided by modified Bessel functions, and when $\nu \to \infty$, the $ \textrm{M}_{\nu} $ covariance function tends to SE kernel. Among other possibilities, $\nu = 7/2$ and $\nu = 9/2 $ values are of primary interest, since these correspond to smooth functions with high predictability of derivatives of higher order, although these are not very suitable for predicting rapid variations. These Matern functions for GP in cosmology were first introduced in \cite{GP_02}. On the other hand, the hyperparameters ${\Theta} \equiv \{\sigma_f, l\}$ are learned by optimising the log marginal likelihood, which is defined as

\begin{equation}
\label{eqn:LML}
\mathcal{L}({\Theta}) = -\frac{1}{2}\textbf{\textrm{y}}^{\textrm{T}}K_{\textrm{y}}^{-1}\textbf{\textrm{y}} -\frac{1}{2}\ln|K_\textrm{y}|+\frac{n}{2}\ln(2 \pi),
\end{equation}
where $K_{\textrm{y}} = K(\textbf{x},\textbf{x}') + C$, $K(\textbf{x},\textbf{x}')$ is the covariance matrix with components $k(x_i,x_j)$, \textbf{y} is the vector of data, $C$ is the covariance matrix of the data for a set of $n$ observations, assuming mean $\mu = 0$. After optimizing for $\sigma_f$ and $l$, one can predict the mean and variance of the function $f(\textbf{x}^{\ast})$ at chosen points $\textbf{x}^{\ast}$ through

\begin{equation}
\label{eqn:MaV}
\begin{split}
< f(\textbf{x}^{\ast}) > &= K(\textbf{x}^{\ast},\textbf{x}) K_{\textrm{y}}^{-1}\textbf{\textrm{y}} \\
\textrm{cov}[f(\textbf{x}^{\ast})] &= K(\textbf{x}^{\ast},\textbf{x}^{\ast}) - K(\textbf{x}^{\ast},\textbf{x}) K_{\textrm{y}}^{-1}K(\textbf{x},\textbf{x}^{\ast}).
\end{split}
\end{equation}

The GP predictions can also be extended to the derivatives of the functions $f(x_i)$,  although limited by the differentiability of the chosen kernel. The derivative of a GP would also be a GP. Thus, one can obtain the covariance between the function and/or the derivatives involved by differentiating the covariance function as

\begin{equation}
    \label{eqn:dcov}
\begin{split}
    \textrm{cov} \left[ f(x_{i}),\dfrac{\partial f(x_{j})}{\partial x_{j}} \right] &= \dfrac{\partial k(x_{i},x_{j})}{\partial x_{j}}  \\
    \textrm{cov} \left[ \dfrac{\partial f(x_{i})}{\partial x_{i}}, \dfrac{\partial f(x_{j})}{\partial x_{j}} \right] &= \dfrac{\partial^{2} k(x_{i},x_{j})}{\partial x_{i}\partial x_{j}}.
\end{split}
\end{equation}

Then, we can write

\begin{equation}
\label{eqn:GP}
f'(x_i)=\mathcal{GP} \left( \mu'(x_i),\textrm{cov} \left[ \dfrac{\partial f(x_{i})}{\partial x_{i}}, \dfrac{\partial f(x_{j})}{\partial x_{j}} \right] \right),
\end{equation}
where $f'(x_i)$ represent the derivatives with respect to their corresponding independent variables, which for our purpose can be the redshift $z$. This procedure can similarly be extended for higher derivatives ($f'(x), f''(x), ..$) in combination with $f(x)$. The mean of the $i^{th}$ derivative and the covariance between $i^{th}$ and $j^{th}$ derivatives, are given by
\begin{equation}
< f^{(i)}(\textbf{x}^{\ast}) > = K^{(i)}(\textbf{x}^{\ast},\textbf{x}) K_{\textrm{y}}^{-1}\textbf{\textrm{y}} 
\end{equation}

\begin{equation}
\label{eqn:dMaV}
\begin{split}
\textrm{cov}[f^{(i)}(\textbf{x}^{\ast}),f^{(j)}(\textbf{x}^{\ast})] &= K^{(i,j)}(\textbf{x}^{\ast},\textbf{x}^{\ast})\\ 
& - K^{(i)}(\textbf{x}^{\ast},\textbf{x}) K_{\textrm{y}}^{-1}K^{(j)}(\textbf{x},\textbf{x}^{\ast}).
\end{split}
\end{equation}

If $i =j$, then we get the variance of the $i^{th}$ derivative in Eq. (\ref{eqn:dMaV}). If the data for derivative functions are available, we can perform a joint analysis, which is the case in our study. Since one data type can be in terms of $f(x)$ while another can be rewritten in terms of $f'(x)$, these different data sets can be combined. In what follows, we describe the data sets that we use in this work.

\begin{figure*}
\begin{center}
\includegraphics[width=3.in]{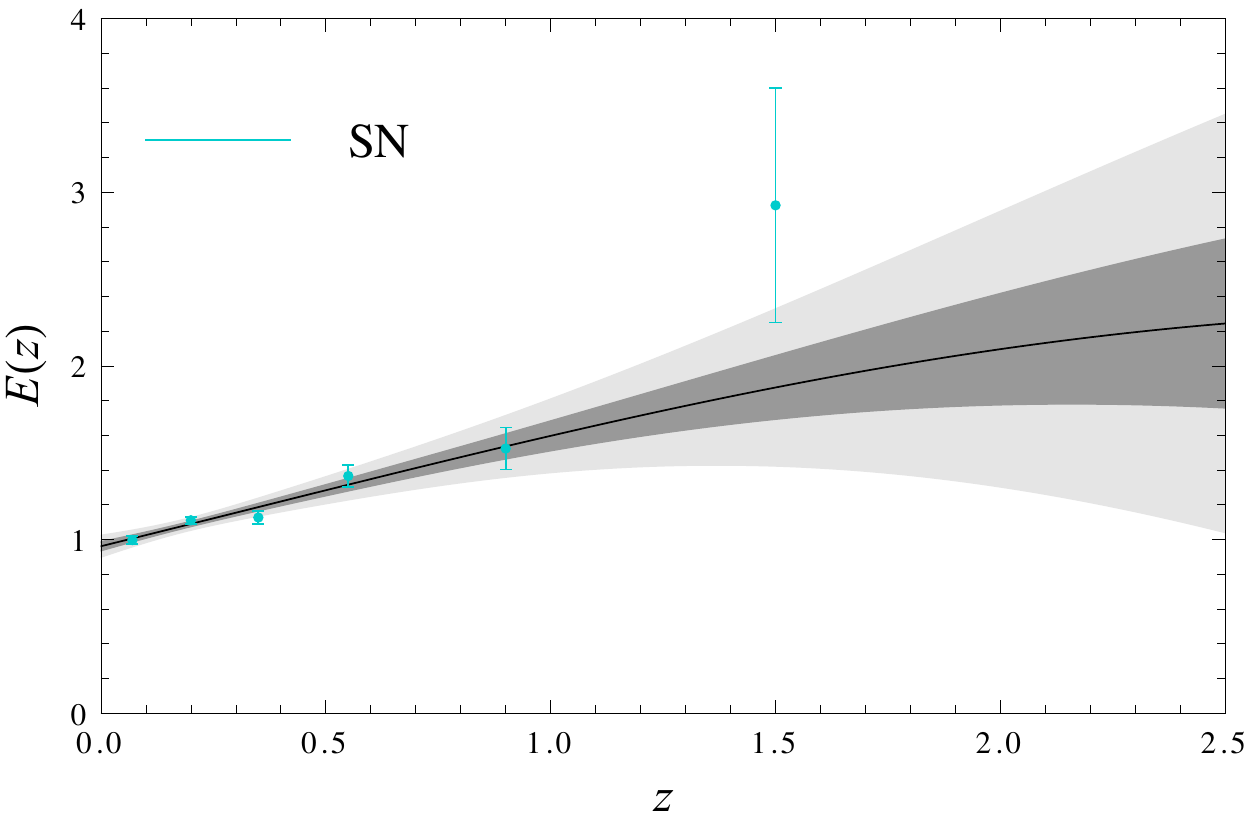} \,\,\,\,\,\,
\includegraphics[width=3.in]{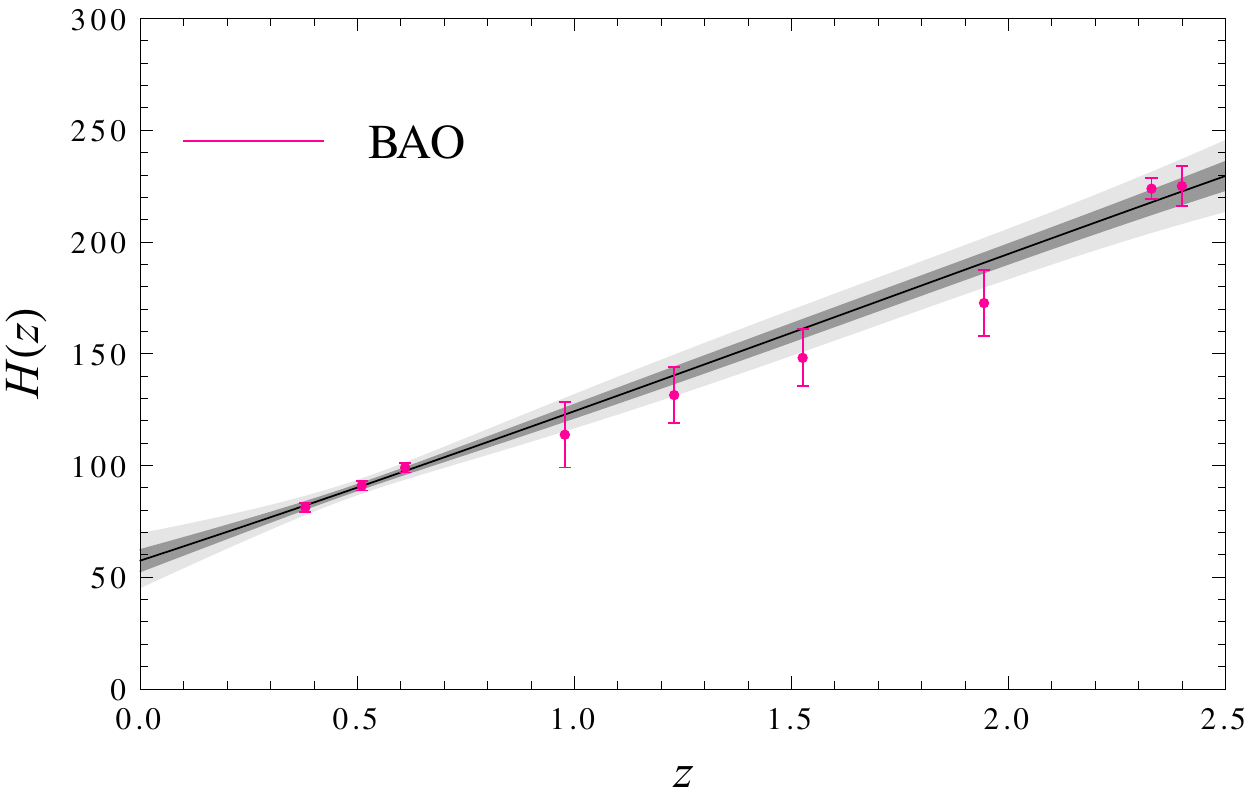}\\
\includegraphics[width=3.in]{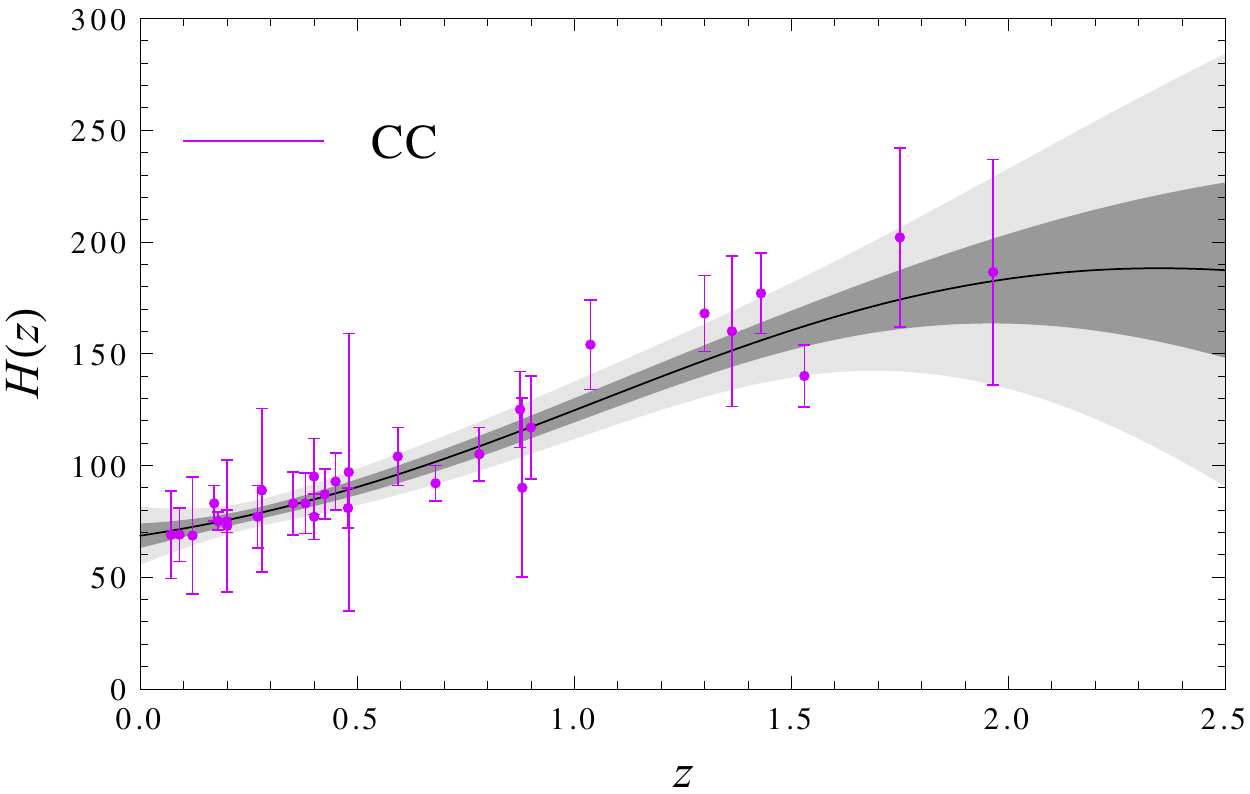} \,\,\,\,\,\,
\includegraphics[width=3.in]{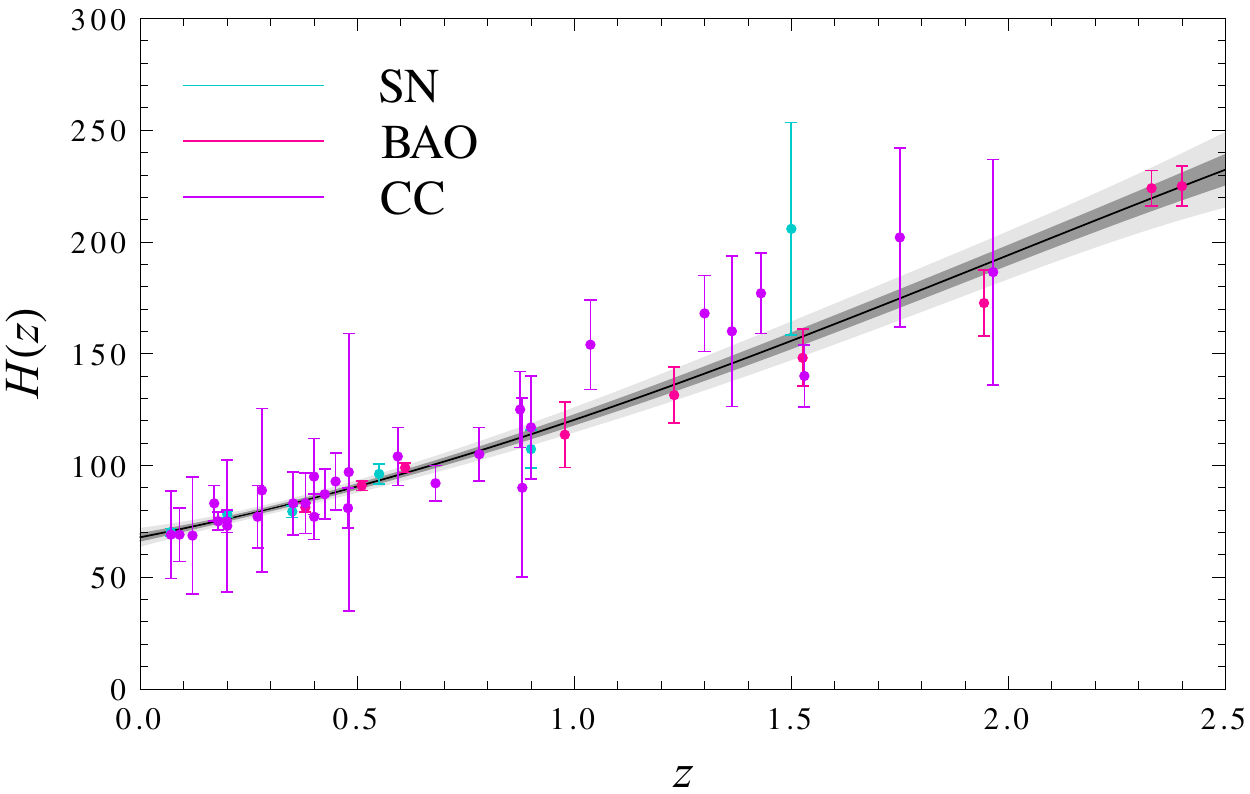}
\caption{$H(z)$ (in units of km s$^{-1}$Mpc$^{-1}$) vs $z$ ($E(z)$ vs $z$ in case of SN data alone) with $1\sigma$ and $2\sigma$ CL regions, reconstructed from SN (top-left),  BAO (top-right), CC (bottom-left) and SN+BAO+CC data (bottom-right). Data with errorbars in all the panels are the observational data as mentioned in the legend of each panel.}
\label{CC_BAO_SN_GP}
\end{center}
\end{figure*}

%$H(z)$ (in units of km s$^{-1}$Mpc$^{-1}$) vs $z$ ($E(z)$ vs $z$ in case of SN data alone) with $1\sigma$ and $2\sigma$ CL regions, reconstructed from SN (top-left),  BAO (top-right), CC (bottom-left) and SN+BAO+CC data (bottom-right), wherein we have used $H_0=68.54 \pm 5.06$ km s$^{-1}$Mpc$^{-1}$ (obtained from the CC analysis) for rescaling the SN data to carry out the joint analysis with SN+BAO+CC data in the bottom-right panel. Data with errorbars in all the panels are the observational data as mentioned in the legend of each panel.

\subsection{Data Sets}

We summarize below the data sets used in our analysis.\\

\textbf{Cosmic Chronometers (CC)}: The CC approach is a powerful method to trace the history of cosmic expansion through the measurement of $H(z)$. We consider the compilation of Hubble parameter measurements provided by \cite{Moresco16}. This compilation consists of 30 measurements distributed over a redshift range $0 <  z <  2$.
\\

\textbf{Baryon Acoustic Oscillations (BAO)}: The BAO is another important cosmological probe, which can trace expanding spherical wave of baryonic perturbations from acoustic oscillations at recombination time through the large-scale structure correlation function, which displays a peak around 150$h^{-1} {\rm Mpc}$. We use BAO measurements from Sloan Digital Sky Survey (SDSS) III DR-12 at three effective binned redshifts $z =$ 0.38, 0.51 and 0.61, reported in \cite{Alam17}, the clustering of the SDSS-IV extended Baryon Oscillation Spectroscopic Survey DR14 quasar sample at four effective binned redshifts $z =$ 0.98, 1.23, 1,52 and 1.94, reported in \cite{Zhao19}, and the high-redshift Lyman-$\alpha$ measurements  at $z = 2.33$ and $z = 2.4$ reported in \cite{du_Mas20} and \cite{du_Mas17}, respectively. Note that the observations are presented in terms of $H(z) \times (r_d/r_{d,fid})$ km s$^{-1}$Mpc$^{-1}$, where $r_d$ is co-moving sound horizon and $r_{d,fid}$ is the fiducial input value provided in the above references. In appendix A, we show that different $r_d$ input values obtained from different data sets do not affect the GP analysis.
%In all the analyses presented in this article, we choose  $r_d/r_{d,fid} = 1$. In appendix A, we show that different $r_d$ input values obtained from different data sets do not affect the GP analysis.
\\

\textbf{Supernovae Type Ia (SN):} The SN traditionally have been one of the most important astrophysical tools in establishing the so-called standard cosmological model. For the present analysis, we use the Pantheon compilation, which consists of 1048 SNIa distributed in a redshift  range $0.01 <  z <  2.3$ \cite{Scolnic18}. Under the consideration of a spatially flat Universe, the full sample of Pantheon can be summarized into six model independent $E(z)^{-1}$ data points \cite{Riess18}. We consider the six data points reported by \cite{Haridasu18} in the form of $E(z)$, including theoretical and statistical considerations made by the authors there for its implementation.
\\

\textbf{H0LiCOW sample:} The Lenses in COSMOGRAIL's Wellspring program\footnote{\url{www.h0licow.org}} have measured six lens systems, making use of the measurements of time-delay distances between multiple images of strong gravitational lens systems by elliptical galaxies \cite{H0LiCOW}. In the analyses of this work, we implement these six systems of strongly lensed quasars reported by the H0LiCOW Collaboration. Full information is contained in the so-called time-delay distance $D_{\Delta t}$. However, additional information can be found in the angular diameter distance to the lens $D_l$, which offers the possibility of using four additional data points in our analysis. Thus, our total H0LiCOW sample  comprises of 10 data points: 6 measurements of time-delay distances and 4 angular diameter distances to the lens for 4 specific objects in the subset information in H0LiCOW sample (see \cite{Birrer2019,Pandey2020} for the description).

\begin{figure*}
\includegraphics[width=3.in]{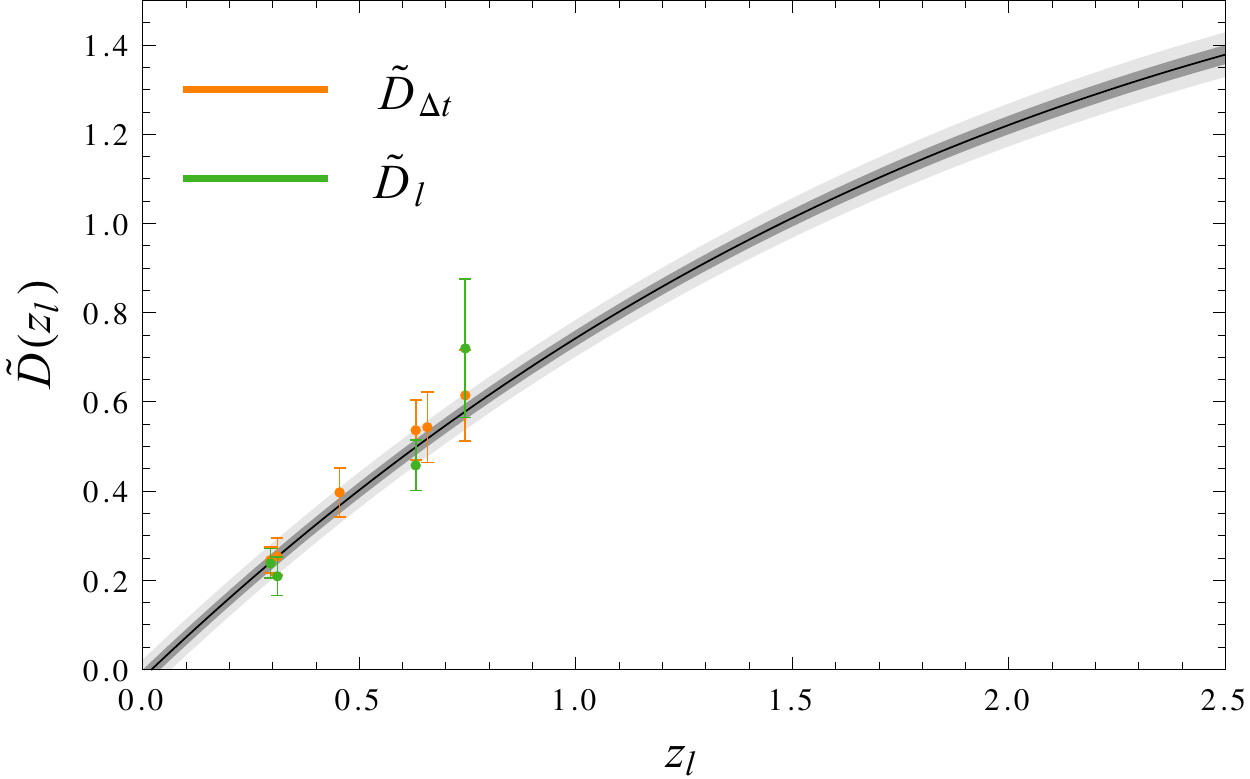} \,\,\,\,\,\,
\includegraphics[width=3.in]{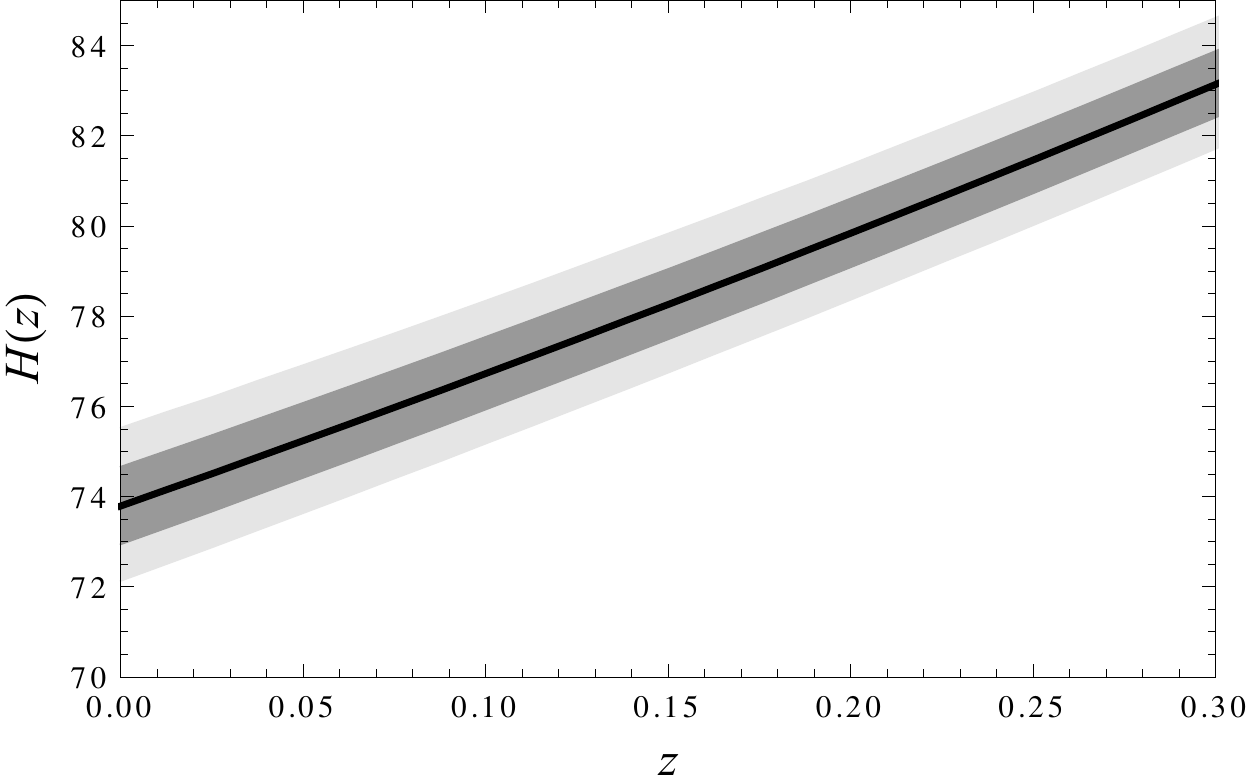}
\caption {Left panel: $\tilde{D}(z_l)$ vs $z_l$ with $1\sigma$ and $2\sigma$ CL regions, reconstructed from H0LiCOW sample data plus other data (SN+CC+BAO). Right panel: $H(z)$ vs $z$  with $1\sigma$ and $2\sigma$ CL regions, enlarged in the redshift range $z<0.3$, and  reconstructed from the combined data SN + BAO + CC + H0LiCOW.}
\label{H0LiCOW_GP_1}
\end{figure*}

\section{Results and Discussions}
\label{results}

First, we verify that analyses carried out from $k_{M_{\nu}}(\tau)$, with $\tau = 9/2$ and $\tau = 7/2$, and $k_{SE}$  do not generate significantly different results, in the sense that all results are compatible with each other at 1$\sigma$ CL, and hence not generating any disagreement/tension between these input kernels. Thus, in what follows, we use GP formalism with an assumed $M_{9/2}$ kernel in the whole analysis. For this purpose, we have used some numerical routines available in the public GaPP code \cite{GP_01}.\\ 

Figure \ref{CC_BAO_SN_GP} shows the reconstructions from SN, BAO and CC data sets, using GP formalism on each data set individually. On the bottom-right panel, we show the $H(z)$ reconstruction from all these data together. First, from the CC reconstruction, we obtain $H_0=68.54 \pm 5.06$ km s$^{-1}$Mpc$^{-1}$, which has been used in the rescaling process of SN data to carry out the joint analysis with SN+BAO+CC (bottom-right). From SN+BAO+CC analysis, we find $H_0=67.85 \pm 1.53$ km s$^{-1}$Mpc$^{-1}$.  Figure \ref{H0LiCOW_GP_1} (left panel) shows the GP reconstruction of $D(z_l)$ from H0LiCOW data, where $z_l$ is the redshift to the lens. On the right panel, we show the reconstruction of $H(z)$ function from SN+BAO+CC+H0LiCOW. In this joint analysis, we obtain $H_0= 73.78 \pm 0.84$ km s$^{-1}$Mpc$^{-1}$, which represents a $1.1\%$ precision measurement. The strategy that we followed to obtain these results is as follows:

\begin{enumerate}
\item The SN+BAO+CC data set is used as in the previous joint analysis, i.e., in terms of $H(z)$ data reconstruction. Thus, now, we just need to re-scale the H0LiCOW data in some convenient way to combine all data for a joint analysis.
    
\item The time-delay distance in H0LiCOW sample is quantified as
    
\begin{equation}
\label{eqn:Dt}
D_{\Delta t} = (1 + z_l) \frac{D_l D_s}{D_{ls}},
\end{equation}
which is a combination of three angular diameter distances, namely $D_l$, $D_s$ and $D_{ls}$, where the subscripts stand for diameter distances to the lens $l$, to the source $s$, and between the lens and the source $ls$.
    
\item At this point, we can get the dimensionless co-moving distance through the relationship 

\begin{equation}
\label{eqn:D}
\tilde{D}(z) = \frac{H_0}{c}(1+z)D_A, 
\end{equation}
where $D_A$ is the angular diameter distance and $\tilde{D}(z)$ is defined as $\tilde{D}(z) = \int^{z}_0 \frac{dz'}{E(z')}$. In this way, we can have: 6 data points from time delay distance $D_{\Delta t}$, which we referred to as $\tilde{D}_{\Delta t}$, and 4 data points obtained from angular diameter distance $D_l$, named as $\tilde{D}_l$. Thus, we can add these 10 data points for joint analysis, and name simply the H0LiCOW sample (see left panel of Figure \ref{H0LiCOW_GP_1}). Note that, to get $\tilde{D}_l$, we directly use the eq. (\ref{eqn:D}), where $D_A = D_l$. On the other hand, to obtain $\tilde{D}_{\Delta t}$, we have to take into account that eq. (\ref{eqn:Dt}) depends on the expansion rate of the Universe through $D_s(z_s,H_0,\Omega_m)$ and $D_{ls}(z_l, H_0,\Omega_m)$, and in this case, we use the $H_0$ and $\Omega_m$ best fit from our SN+BAO+CC joint analysis. 

\item For the joint analysis, the relation $\tilde{D}(z) = \int^{z}_0 \frac{dz'}{E(z')}$ can be reversed to obtain $E(z) = \frac{1}{\tilde{D}'(z)}$. So, we can make use of this possibility that offers the reconstruction of the first derivative of the dimensionless co-moving distance $\tilde{D}'(z)$. For this purpose, we introduce the SN + BAO + CC  data set in the form of $1/E(z)$ and the H0LiCOW data set in the form of $\tilde{D}(z)$, to obtain the GP reconstruction of dimensionless co-moving distance.
\end{enumerate}

\begin{figure}
%\begin{center}
\includegraphics[width=3.4in]{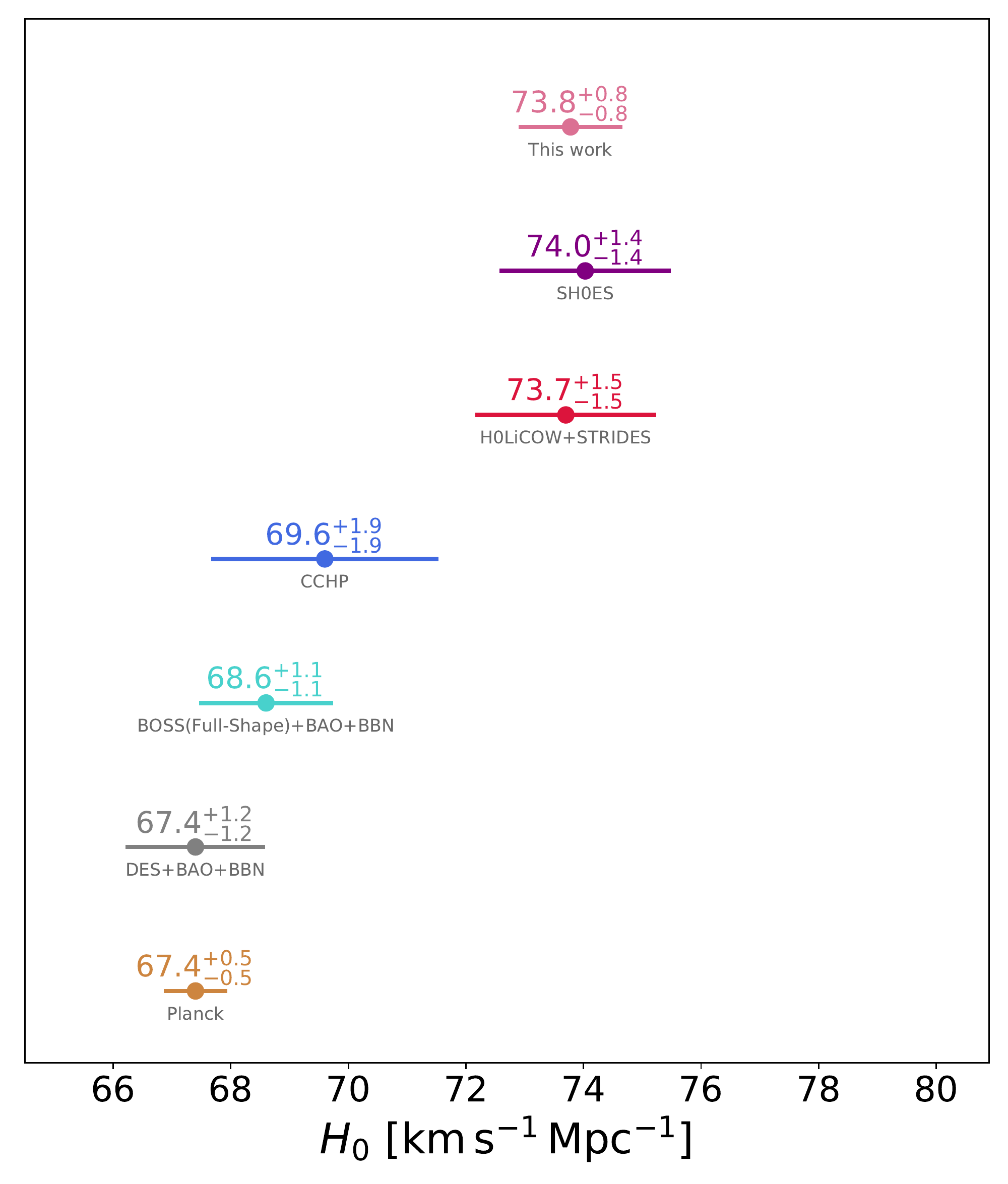} 
\caption{Compilation of $H_0$ measurements taken from recent literature, namely, from Planck collaboration (Planck) \cite{Planck2018}, Dark Energy Survey Year 1 Results (DES+BAO+BBN) \cite{DES}, the final data release of the BOSS data (BOSS Full-Shape+BAO+BBN) \cite{BOSS_H0}, The Carnegie-Chicago Hubble Program (CCHP) \cite{Freedman}, H0LiCOW collaboration (H0LiCOW+STRIDES) \cite{H0LiCOW}, SH0ES \cite{R19}, in comparison  with the $H_0$ constraints obtained in this work from the GP analysis using SN+BAO+CC+H0LiCOW.}
\label{H0_GP}
%\end{center}
\end{figure}

From the joint analysis SN+BAO+CC+H0LiCOW, we find $H(z=0)= 73.78 \pm 0.84$ km s$^{-1}$Mpc$^{-1}$. Figure \ref{H0LiCOW_GP_1} (right panel) shows the $H(z)$ reconstruction from SN+BAO+CC+H0LiCOW. Figure \ref{H0_GP} shows a comparison of our joint analysis estimates on $H_0$ with others recently obtained in literature. We note that our constraint on $H_0$ is in accordance with SH0ES and H0LiCOW+STRIDES estimates. On the other hand, we find $\sim$6$\sigma$ tension with current Planck-CMB measurements and $\sim$2$\sigma$ tension with CCHP best fit. We re-analyze our estimates removing BAO data (see appendix A). In this case, we find $H_0 = 68.57 \pm 1.86$ km s$^{-1}$Mpc$^{-1}$ and $H_0 = 71.65 \pm 1.09 $ km s$^{-1}$Mpc$^{-1}$ from SN+CC and SN+CC+H0LiCOW, respectively.
\\

\begin{figure*}
\begin{center}
\includegraphics[width=3.in]{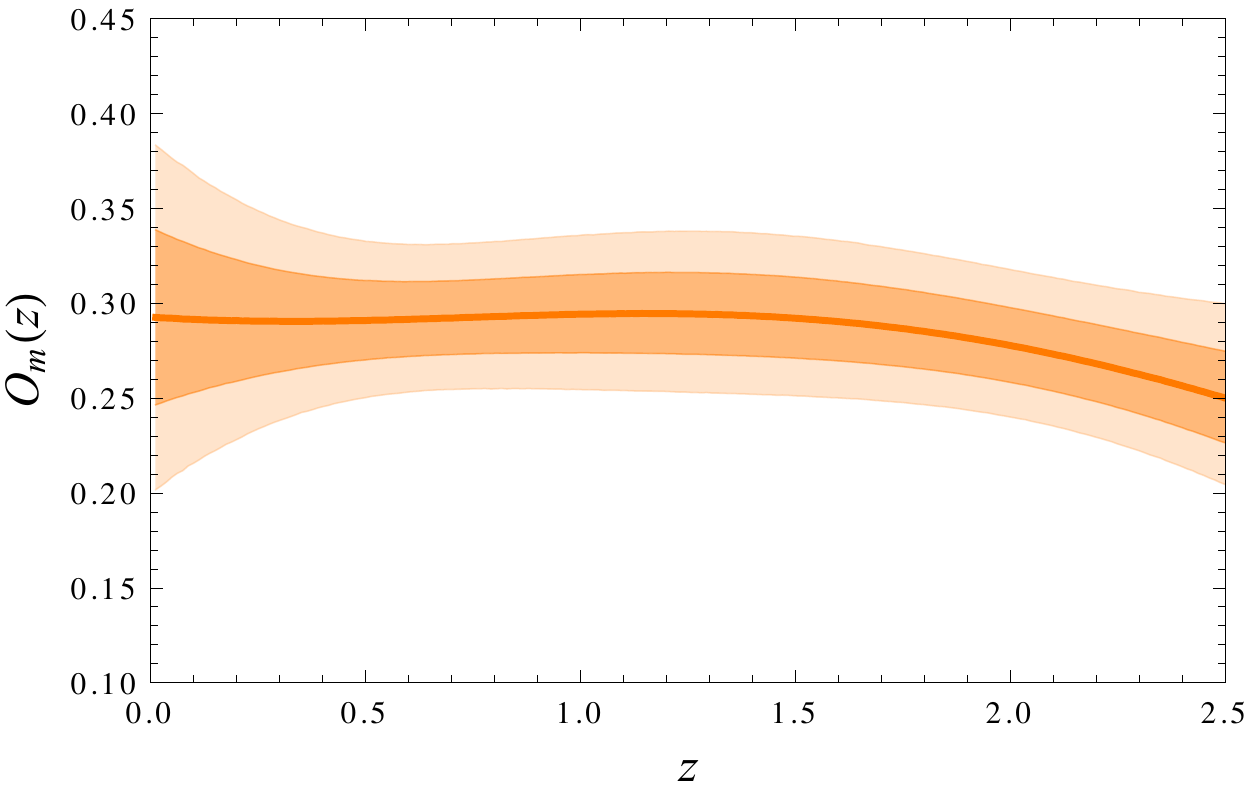} \,\,\,\,
\includegraphics[width=3.in]{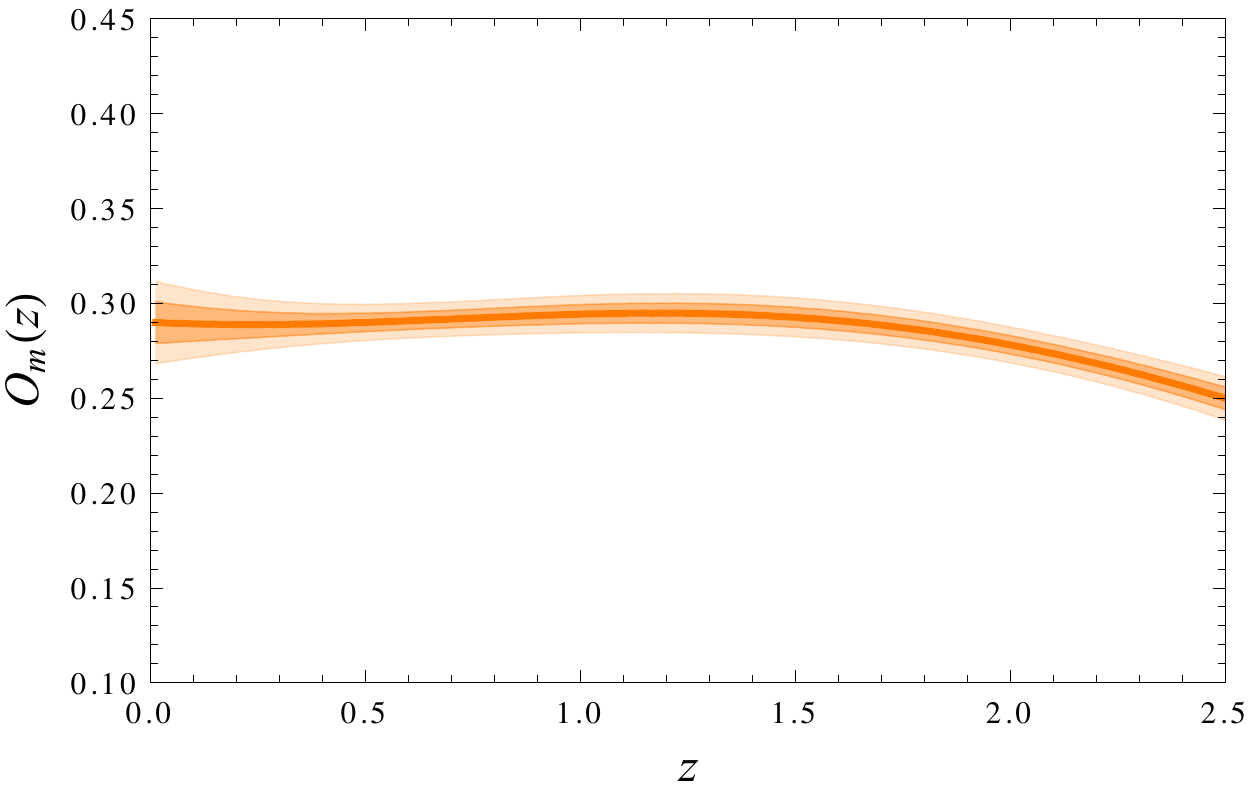}
\caption{$O_m(z)$ vs $z$  with $1\sigma$ and $2\sigma$ CL regions, reconstructed from SN+BAO+CC data (left panel) and SN+BAO+CC+H0LiCOW data (right panel).}
\label{Om_results}
\end{center}
\end{figure*}

In the context of the standard framework, we can also check the $O_m(z)$ diagnostic \cite{Omz_01}

\begin{equation}
\label{eqn:Omz}
O_m(z) = \frac{E^2(z) - 1}{(1 + z)^3 - 1}.
\end{equation}

If the expansion history $E(z)$ is driven by the standard $\Lambda$CDM model, then $O_m(z)$ is practically constant and equal to the density of matter $\Omega_{m}$, and so, any deviation from this constant can be used to infer the dynamical nature of DE. Figure \ref{Om_results} shows the reconstruction of the $O_m(z)$ diagnostic. We find $\Omega_m = 0.292 \pm 0.046$ and $\Omega_m = 0.289 \pm 0.012$ at $1\sigma$ from  SN+BAO+CC and SN+BAO+CC+H0LiCOW analyses, respectively. To obtain these results, we normalize $H(z)$ with respect to $H_0$ to obtain $E(z)$ for the entire data set except SN, where $H_0$ is taken from SN+BAO+CC, and SN+BAO+CC+H0LiCOW cases, respectively. The prediction from SN+BAO+CC is compatible with $\Omega_m = 0.30$ across the analyzed range, but it is interesting to note that for $z > 2$, we have $\Omega_m < 0.30$ at $\sim$2$\sigma$ from SN+BAO+CC+H0LiCOW. These model-independent $\Omega_m$ estimates will be used as input values in the reconstruction of $w$.

\begin{figure*}
\begin{center}
\includegraphics[width=3.in]{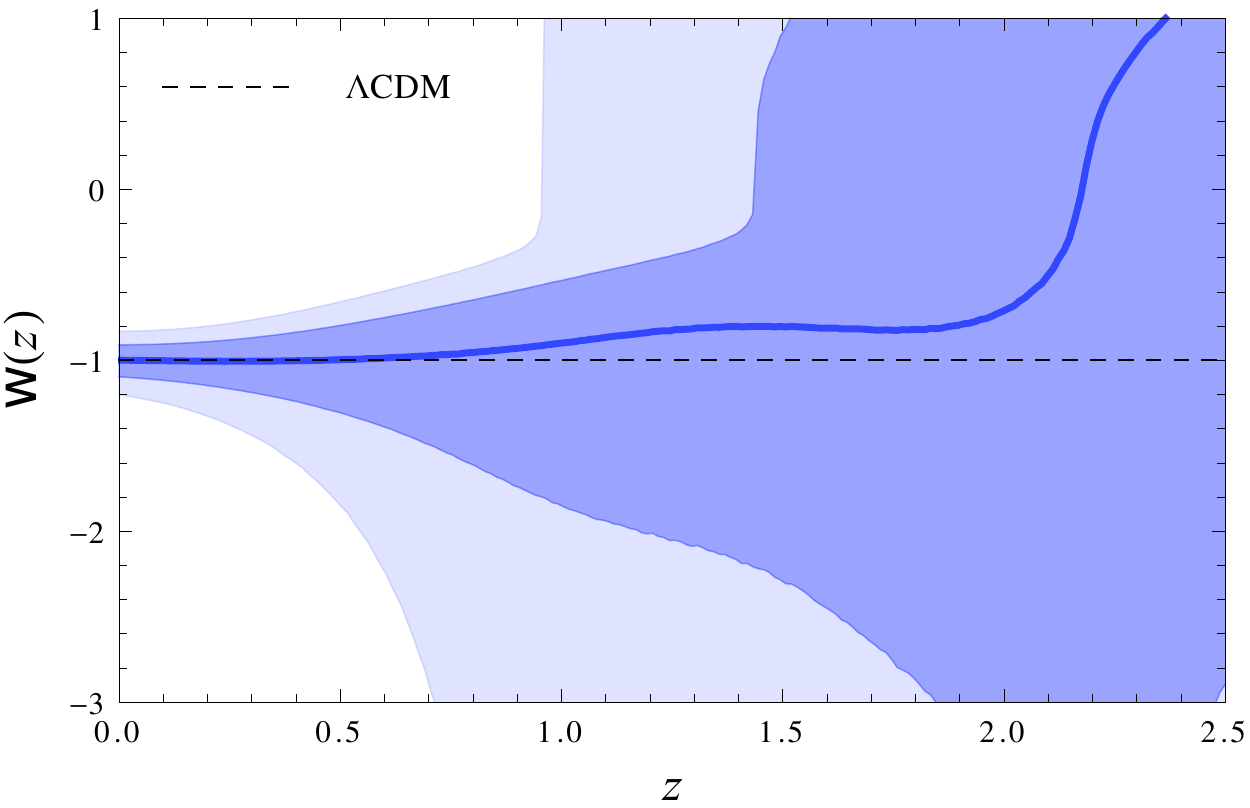} \,\,\,\,
\includegraphics[width=3.in]{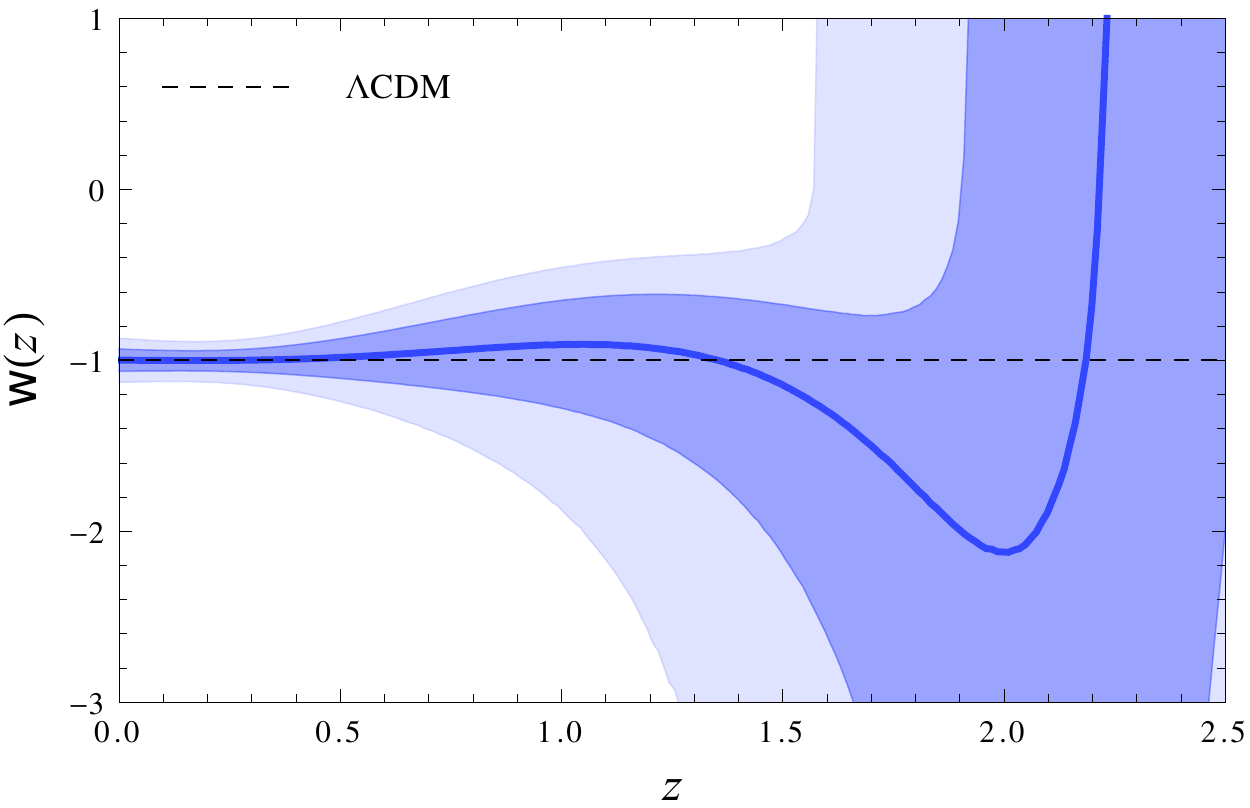}
\caption{The EoS $w(z)$ vs $z$  with $1\sigma$ and $2\sigma$ CL regions, reconstructed from SN+BAO+CC data (left panel) and SN+BAO+CC+H0LiCOW data (right panel).}
\label{w_results}
\end{center}
\end{figure*}

\begin{figure*}
\begin{center}
\includegraphics[width=3.in]{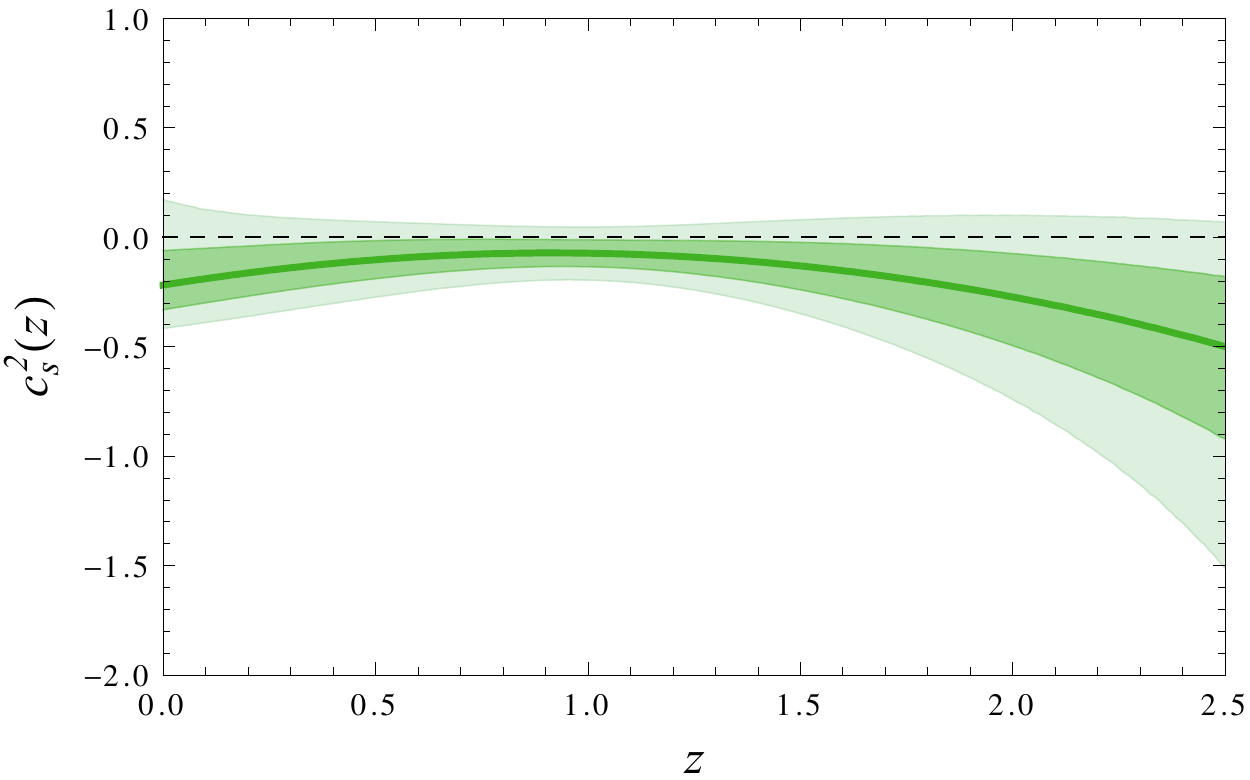} \,\,\,\,
\includegraphics[width=3.in]{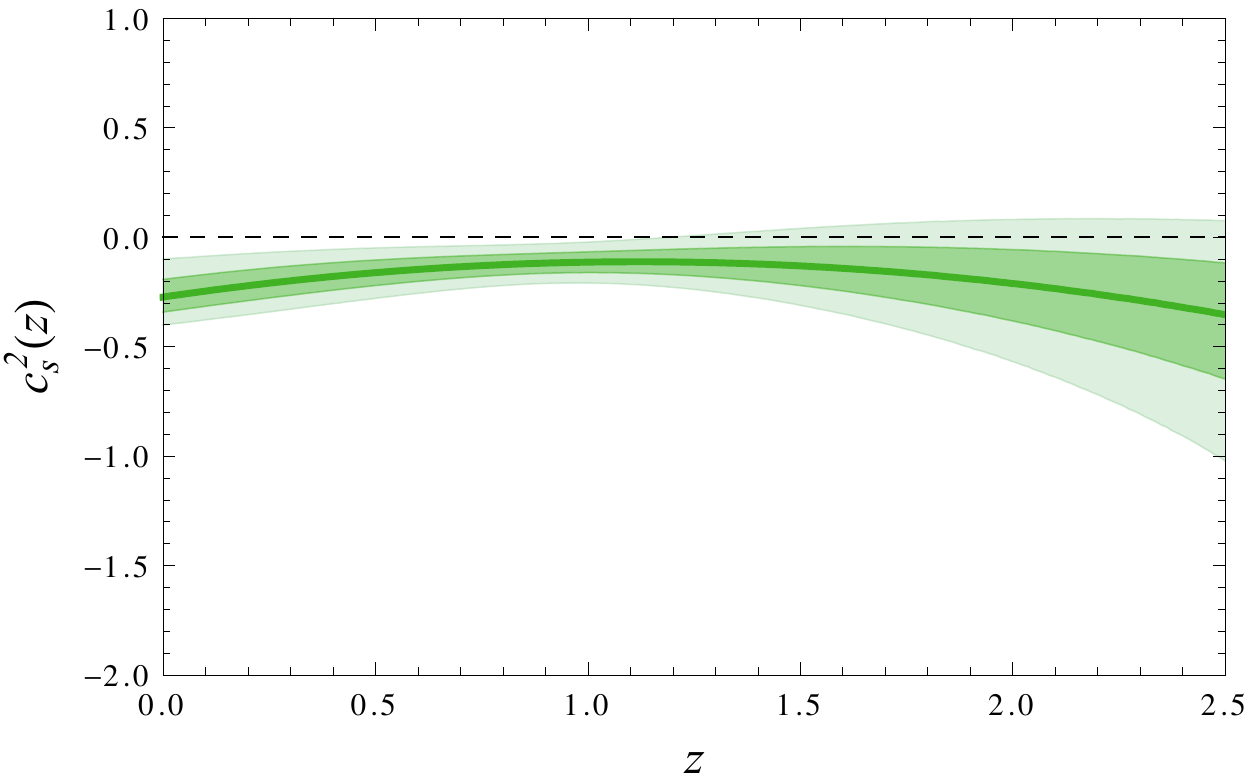}
\caption{$c_s^2(z)$ vs $z$  with $1\sigma$ and $2\sigma$ CL regions, reconstructed from SN+BAO+CC data (left panel) and SN+BAO+CC+H0LiCOW data (right panel). The dashed line is the division of dark energy clustering.}
\label{c2s_results}
\end{center}
\end{figure*}

\begin{figure*}
\begin{center}
\includegraphics[width=3.2in]{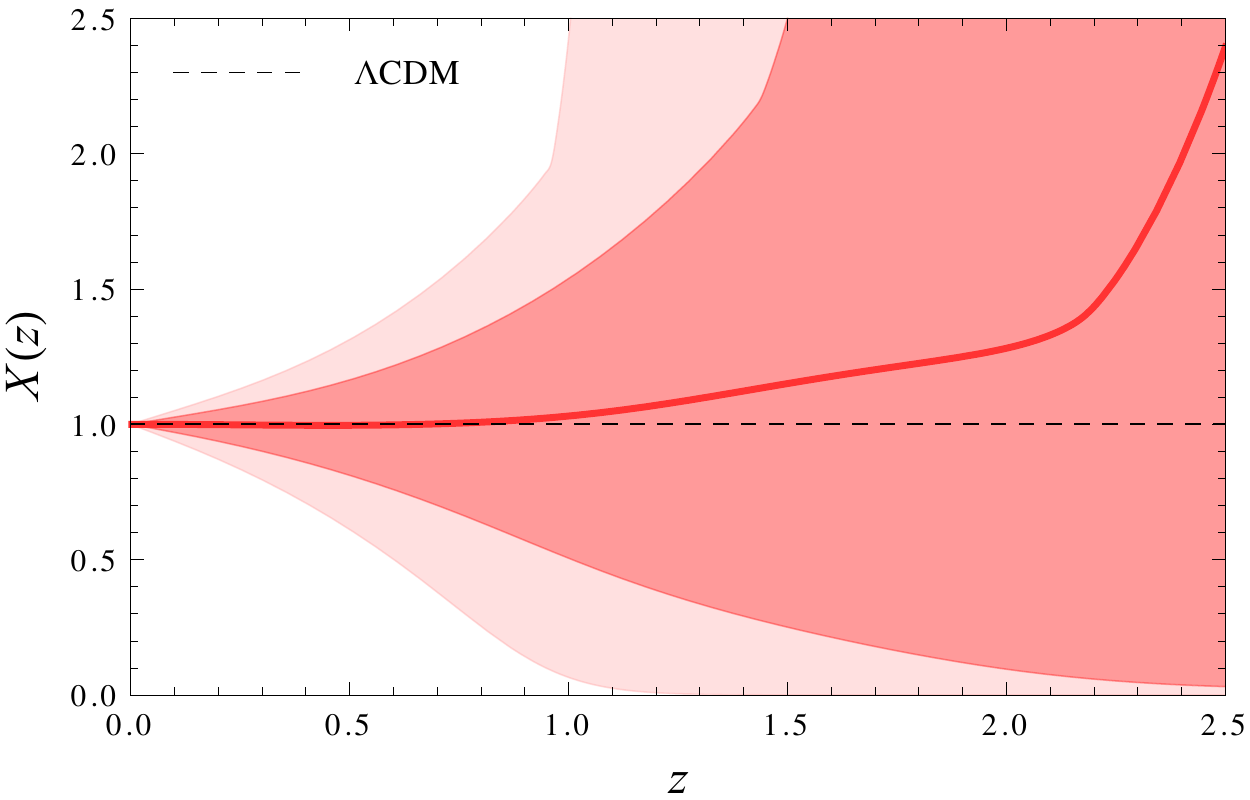} \,\,\,\,
\includegraphics[width=3.2in]{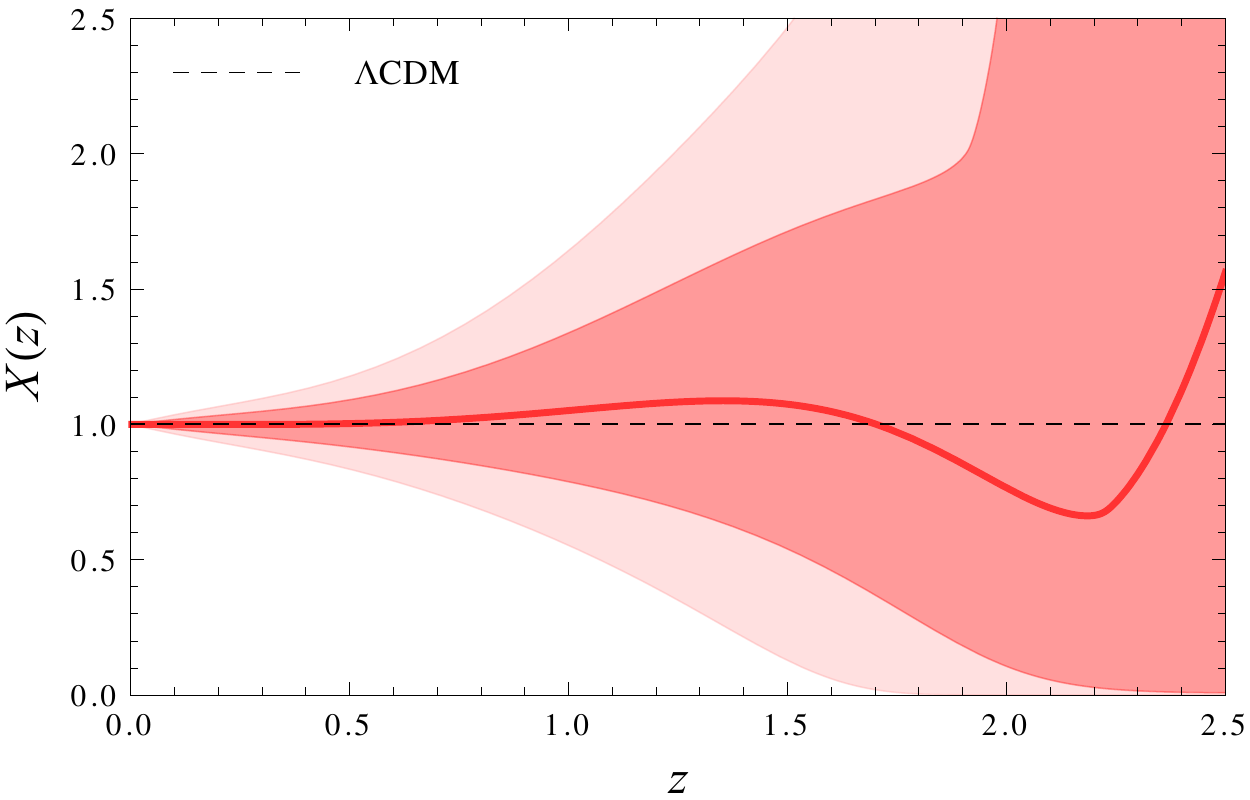}
\caption{$X(z)$ vs $z$  with $1\sigma$ and $2\sigma$ CL regions, reconstructed from SN+BAO+CC data (left panel) and SN+BAO+CC+H0LiCOW data (right panel). The dashed black curve corresponds to $\Lambda$CDM model prediction $X = 1$.}
\label{Xz_results}
\end{center}
\end{figure*}

The EoS of DE can be written as \cite{w_01,w_02,w_03}
\begin{equation}
\label{eqn:EoS}
\begin{split}
w(z) &= \frac{2(1 + z)E(z)E'(z) - 3E^2(z) + \Omega_{k} (1 + z)^2}{3\left( E^2(z) - \Omega_{m} (1 + z)^3 - \Omega_{k} (1 + z)^2 \right)},
\end{split}
\end{equation}
where $\Omega_{m}$ and $\Omega_k$ are the density parameters of matter (baryonic matter + dark matter) and spatial curvature, respectively. In what follows, we assume $\Omega_k = 0$, which is a strong, though quite general assumption about spatial geometry. 

Figure \ref{w_results} shows the $w(z)$ reconstruction from SN+BAO+CC and SN+BAO+CC+H0LiCOW data combinations on the left and right panels, respectively. From both analyses, we notice that $w$ is well constrained for $z \lesssim 0.5$ with the prediction $w=-1$. Most of the data correspond to this range in numbers and precision. The GP mean excludes any possibility of $w \neq -1$ in the whole range of $z$ under consideration. We observe that the best fit prediction is on $w= -1$ up to $z \sim 0.5$ for both cases. The addition of the H0LiCOW data considerably improve  the reconstruction of $w$ for $z < 1$. Beyond this range, the best fit prediction can deviate from $w = -1$, but statistically compatible with a cosmological constant. Evaluating at the present moment, we find $w(z=0)= -0.999 \pm 0.093$ and $w(z=0)= -0.998 \pm 0.064$ from SN+BAO+CC and SN+BAO+CC+H0LiCOW, respectively. Note that H0LiCOW sample improves  the constraints on $w(z=0)$ up to $\sim$2.9\%.
\\

From the statistical reconstruction of $w(z)$ and its derivative $w'(z)$, we can analyze the DE adiabatic sound speed $c^2_s$. Given the relation $p=w\rho$, we can find 

\begin{equation}
c^2_s (z) = \frac{\delta p}{\delta \rho} = w(z) + \frac{1 + z}{3} \frac{w'(z)}{1 + w(z)}.
\end{equation}

Figure \ref{c2s_results} shows $c^2_s$ reconstruction from SN+BAO+CC and SN+BAO+CC+H0LiCOW data combinations on the left and right panels, respectively. We note that the DE sound speed is negative at $\sim$1$\sigma$ from SN+BAO+CC when evaluated up to $z \simeq 2.5$. It is interesting to note that the SN+BAO+CC+H0LiCOW  analysis yields $c^2_s < 0$ at 2$\sigma$ for $z < 1$. At the present moment, we find $c^2_s(z=0) = -0.218 \pm 0.137$ and $c^2_s(z=0) = -0.273 \pm 0.068$ at 1$\sigma$ CL from SN+BAO+CC and SN+BAO+CC+H0HiCOW, respectively. Therefore, this inference on $c^2_s$ rules out significantly the possibility for clustering DE models, and also the models with $c^2_s > 0$ up to high $z$ at least at 1$\sigma$ CL. The condition $c^2_s > 0$ is usually imposed to avoid gradient instability. However, the perturbations can still remain stable under $c^2_s < 0$ consideration \cite{Arjona_01,Arjona_02,Arjona_03, Cardona}. Thus, if the effective sound speed is negative, this would be a smoking gun signature for the existence of an anisotropic stress and possible modifications of gravity. Recently, a possible evidence for $c^2_s < 0$ is found in \cite{D_8}, and also in a model-independent way from the Hubble data.
Now, we look at some models which can potentially explain this result.

The Lagrangian $L = G_2(\phi, X) + \frac{M^2_{pl}}{2}R$ describes general K-essence scenarios. Here the function $G_2$ depends on $\phi$ and $X = -\frac{1}{2} \nabla^{\mu} \phi \nabla_{\mu} \phi$, and $R$ is the Ricci scalar curvature. In this case, the sound speed is given by
\begin{equation}
\label{K_model}
c^2_s = \frac{G_{2,X}}{G_{2,X} + \dot{\phi} G_{2,XX}},
\end{equation}
where $G_{2,X} \equiv \partial G_2 /\partial X$. Quintessence models correspond to the particular choice $G_2 = X - V(\phi)$, given $c^2_s = 1$. Thus, the usual quintessence scenarios are discarded from our results, which predict negative or low values of the sound speed.

Considering the so-called dilatonic ghost condensate \cite{ghost_condensate}, given by the Lagrangian,
\begin{equation}
G_2 = -X + e^{\lambda \phi/M_{pl}} \frac{X^2}{M^4},
\end{equation}
where $\lambda$ and $M$ are free parameters of the model, we can write $c^2_s$ as
\begin{equation}
c^2_s = \frac{2y -1}{6y-1},
\end{equation}
with $y = \frac{\dot{\phi}^2 e^{\lambda \phi/M_{pl}}}{2M^4}$. The condition $y < -1/2$ ensures negative sound speed values.

Another interesting possibility pertains to a unified dark energy and dark matter scenario described by $G_2 =-b_0 + b_2(X-X_0)^2$, where $b_0$ and $b_2$ are free parameters of the model \cite{Scherrer}. In this case, the sound speed is
\begin{equation}
c^2_s = \frac{X-X_0}{3X - X_0},
\end{equation}
where $c^2_s < 0$ for $X < X_0$.

The above mentioned cases are theoretical examples under the consideration of a minimally coupled gravity scenario, which can reproduce a possible $c^2_s < 0$ behavior. More generally, in the Horndeski theories of gravity \cite{Horndeski,Deffayet,Kobayashi11}, the speed of sound can be written as
\begin{eqnarray}
\label{c2s}
\alpha c^2_s = \Big[ \Big(1- \frac{\alpha_B}{2} \Big) (2\alpha_M + \alpha_B) + \frac{\alpha_B}{2} (\ln H^2)' + \alpha'_B \Big],
\end{eqnarray}
where prime denotes $d/d\ln a$, and $\alpha_i$ are functions expressed in a way that highlights their effects on the theory space \cite{Bellini}, namely, kineticity ($\alpha_K$), braiding ($\alpha_B$) and Planck-mass run rate ($\alpha_M$). Further, we define $\alpha = \alpha_K + 3/2 \alpha^2_B$. Motivated for the tight constraints on the difference between the speed of gravitational waves  and the speed of light to be $\lesssim 10^{-15}$ from the GW170817 and GRB 170817A observations \cite{Gw07,Gw08}, we assume $\alpha_T = 0$ (tensor speed excess). Without loss of generality, we can consider $\alpha > 0$ and the relation $\alpha_B = R \times \alpha_M$, with $R$ being a constant. For instance, for $R=-1$, we reproduce $f(R)$ gravity theories. Different $R$ values can manifest the most diverse possible changes in gravity. For a qualitative example, taking $R=-1$, the running of the Planck mass must satisfy the relationship
\begin{eqnarray}
\label{}
\frac{3}{2}\alpha^2_M - a \frac{dH}{da}\frac{\alpha_M}{H} - a \frac{d \alpha_M}{da} \leq 0,
\end{eqnarray}
for generating $c^2_s < 0$. At late cosmic time, we have $\frac{dH}{da}\frac{1}{H} < 0$, and we can consider the theories in a good approximation where $|\alpha_M \ll 1|$. So we see that the condition $\alpha_M < 0$, can generate negative $c^2_s$ values in this case.
\\

Finally, we analyze the function
\begin{equation}
X(z) = \frac{\rho_{\rm de}}{\rho_{\rm de,0}} = \exp\left( 3 \int^{z}_0 \frac{1 + w(z')}{1 + z'} dz' \right),
\end{equation}
quantifying the ratio of DE energy density evolution over the cosmic time. 

Figure \ref{Xz_results} shows $X(z)$ reconstruction from SN+BAO+CC and SN+BAO+CC+H0LiCOW data combinations on the left and right panels, respectively. We note that the evolution of $X$ is fully compatible with the $\Lambda$CDM model, and with the best fit model-independent prediction around $X= 1$ up to $z \sim 1$, in both analyses. It is interesting to note that $X$ can cross to negative values when $z > 1$ and $z > 1.5$ at 2$\sigma$ CL from SN+BAO+CC and SN+BAO+CC+H0LiCOW, respectively. It can also have some interesting theoretical consequences. First, DE with negative density values at large $z$ came to the agenda when it turned out that, within the standard $\Lambda$CDM model, the Ly-$\alpha$ forest measurement from BAO data by the BOSS collaboration \cite{X_01}, prefers a smaller value of the dust density parameter compared to the value preferred by the CMB data. Thus, with the possibility of a preference for negative energy density values at high $z$, it is argued that the Ly-$\alpha$ data at $z \sim 2.34$ can be described by a non-monotonic evolution in $H(z)$ function, which is difficult to achieve in any model with non-negative DE density \cite{X_02}. Note that in our analysis, we are taking into account the high $z$ Lyman-$\alpha$  measurements reported in \cite{du_Mas20} and \cite{du_Mas17}. It is possible to achieve $X < 0$ at high $z$ when the cosmological gravitational coupling strength gets weaker with increasing $z$ \cite{X_03,X_04}. A range of other examples of effective sources crossing the energy density below zero also exists, including theories in which the cosmological constant relaxes from a large initial value via an adjustment mechanism \cite{X_05}, and also by modifying gravity theory \cite{X_06,X_07,X_08}. More recently, a graduated DE model characterized by a minimal dynamical deviation from the null inertial mass density is introduced in \cite{X_09}  to obtain negative energy density at high $z$. Also, seeking inspiration from string theory, the possibility of negative energy density is investigated  in \cite{X_10}.

The reconstruction of $w(z)$ and $X(z)$ are robust at low $z$, where the DE effects begin to be considerable, and a slow evolution of the EoS is well captured at 68\% CL. However, the error estimates are larger at high $z$, where the data density is significantly smaller and the dynamical effects of DE are weaker. The introduction of the H0LiCOW data slightly improves the estimated errors in this range, especially for $1.0<z<1.5$. On the other hand, the uncertainties of smooth functions may have a greater amplitude than the highly oscillating functions, and in this way the propagation of errors to their derivatives can be overestimated \cite{Rasmussen}. In our case, the variation of the starting functions is quite smooth with respect to the data and their derivatives as well, leading to the propagation of errors with a greater amplitude, as can be seen in Figures \ref{w_results} and \ref{Xz_results} at high $z$. Other aspects that may influence this fact could be the strong dependence on $z$, as in the case of $w(z)$, and the integrability of the functions with respect to $z$, as in the case of $X(z)$ (for a brief discussion in this regard, see for example \cite{GP_01}). 

Recently, the authors in \cite{Renzi_Silvestri} have obtained a measurement $H_0= 69.5 \pm 1.7$ km s$^{-1}$Mpc$^{-1}$, showing that it is possible to constrain $H_0$ with an accuracy of 2\% with minimal assumptions, from a combination of independent geometric datasets, namely, SN, BAO and CC. They have not used the H0LiCOW data in their analyses as we have used in the present work. They have also reconstructed the DE density parameter $X(z)$, finding similar conclusion as obtained here in this work.

\section{Final remarks}
\label{conclusion}
We have applied GP to constrain $H_0$, and to reconstruct some functions that describe physical properties of DE in a model-independent way using cosmological information from SN, CC, BAO and H0LiCOW lenses data. The main results from the joint analysis, i.e.,  SN+CC+BAO+H0LiCOW, are summarized as follows:
\\

i) A 1.1\% accuracy measurement of $H_0$ is obtained with the best fit value $H_0= 73.78 \pm 0.84$ km s$^{-1}$Mpc$^{-1}$ at 1$\sigma$ CL.
\\

ii) The EoS of DE is measured at $\sim$ 6.5\% accuracy at the present moment, with $w(z=0)=-0.98 \pm 0.064$ at 1$\sigma$ CL.
\\

iii) We find possible evidence for $c^2_s < 0$ at $\sim$ 2$\sigma$ CL from the analysis of the function behavior at high $z$. At the present moment, we find $c^2_s(z=0) = -0.273 \pm 0.068$ at 1$\sigma$ CL.
\\

iv) We find that the ratio of DE density evolution, $\rho_{\rm de}/\rho_{\rm de,0}$, can cross to negative values at high-$z$. This behavior has already been observed by other authors. Here, we re-confirm this possibility for $z > 1.5$ at $\sim$2$\sigma$.  
\\

Certainly, the GP method having the ability to perform joint analysis has a great potential in search for the accurate measurements of cosmological parameters, and analyze physical properties of the dark sector of the Universe in a minimally model-dependent way. It can shed light in the determination of the dynamics of the dark components or even rule out possible theoretical cosmological scenarios. Beyond the scope of the present work, it will be interesting to analyze/reconstruct a possible interaction in the dark sector, where DE and dark matter interact non-gravitationally in a model-independent way, through a robust joint analysis. Such scenarios have been intensively investigated recently in literature. We hope to communicate results in that direction in near future.

\begin{figure*}[hbt!]
\begin{center}
\includegraphics[width=3.in]{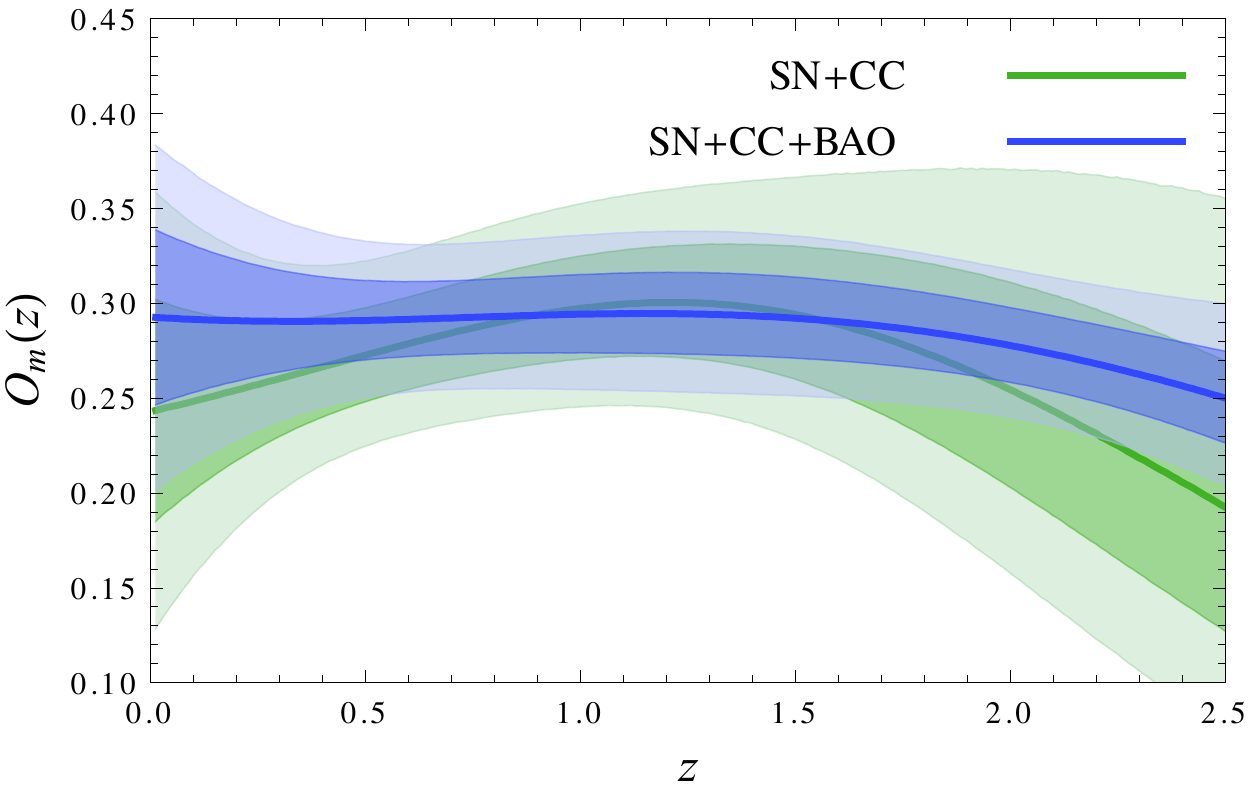} \,\,\,\,
\includegraphics[width=3.in]{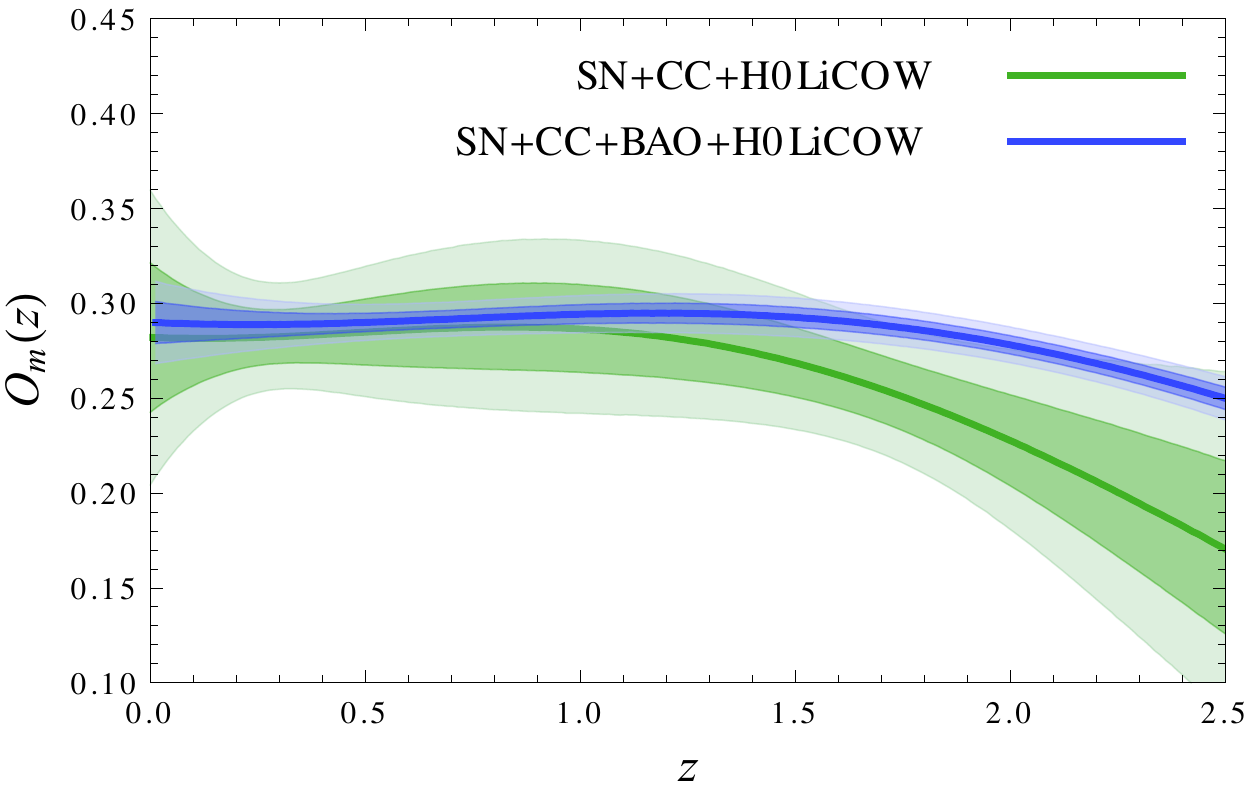}
\caption{$O_m(z)$ vs $z$  with $1\sigma$ and $2\sigma$ CL regions, reconstructed from SN+CC (green) and SN+CC+BAO (blue) in the left panel, and SN+CC+H0LiCOW (green) and SN+CC+H0LiCOW+BAO (blue) in the right panel.}
\label{Om_results_appendix}
\end{center}
\end{figure*}

\begin{figure*}
\begin{center}
\includegraphics[width=3.in]{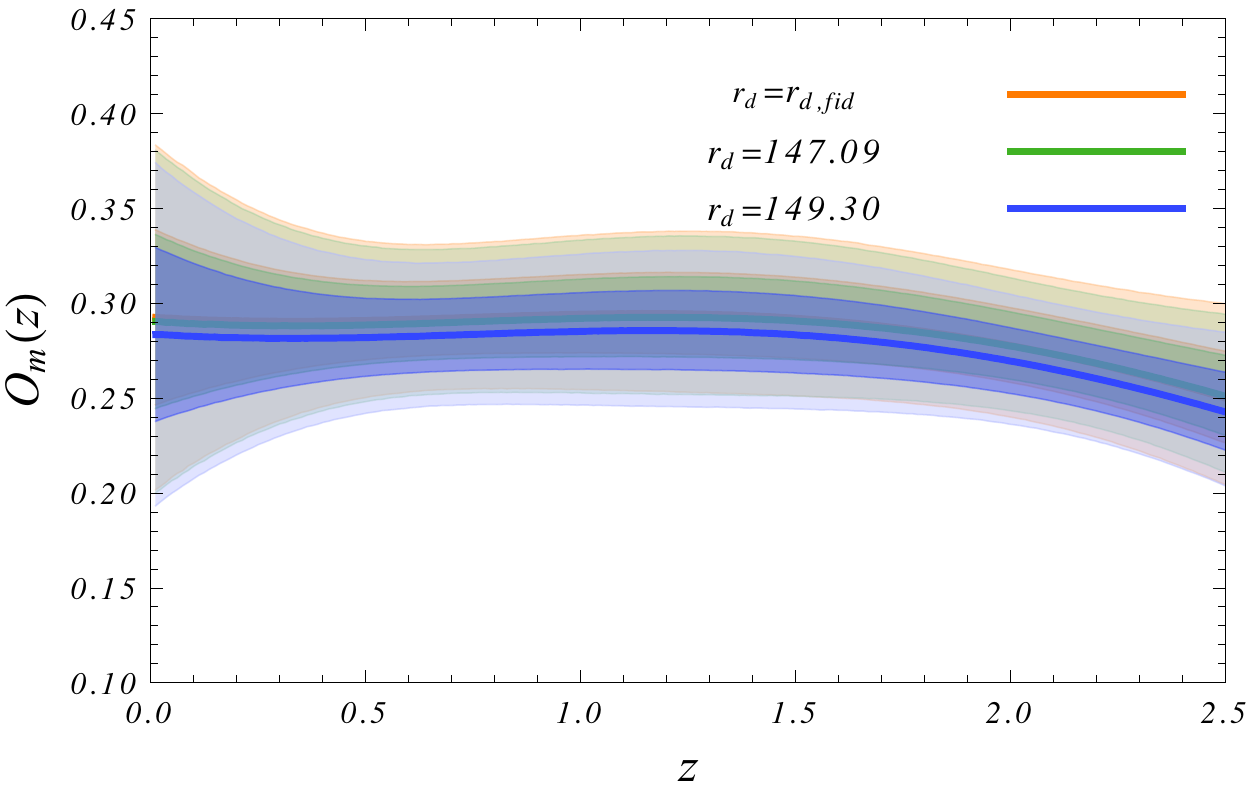} \,\,\,\,
\includegraphics[width=3.in]{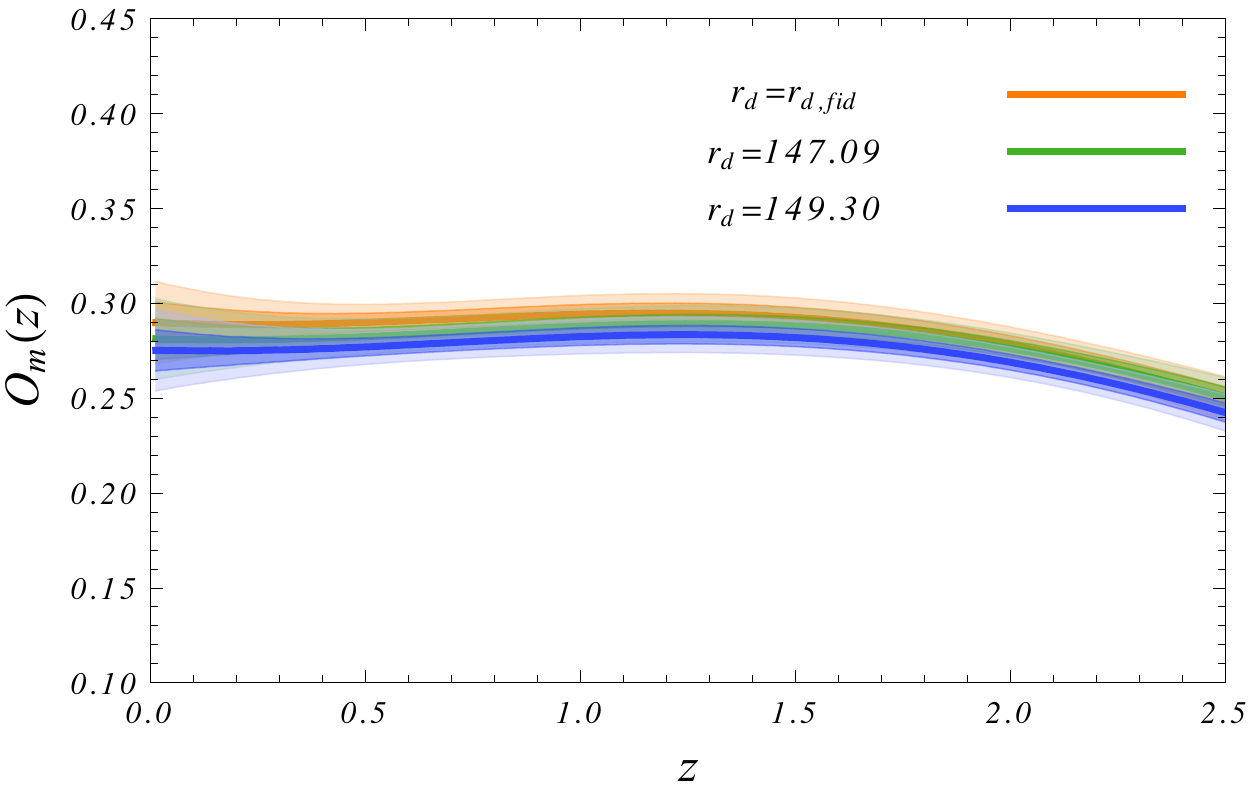}\\
\caption{$O_m(z)$ vs $z$  with $1\sigma$ and $2\sigma$ CL regions for different values of $r_d$ (in units of Mpc), reconstructed from SN+BAO+CC data (left panel) and SN+BAO+CC+H0LiCOW data (right panel).}
\label{Om_results_appendix_2}
\end{center}
\end{figure*}

\appendix
\section{$H_0$ without BAO data, and effects of $r_d$}
In this appendix, we derive constraints on $H_0$ and $O_m(z)$ diagnostic removing our BAO data set compilation as described in section II. Figure \ref{Om_results_appendix} shows $O_m(z)$ vs $z$ reconstructed from SN+CC  and SN+CC+H0LiCOW. For comparison, we also show the prediction with BAO. We find $H_0 = 68.57 \pm 1.86$ km s$^{-1}$Mpc$^{-1}$ and $H_0 = 71.65 \pm 1.09$ km s$^{-1}$Mpc$^{-1}$ from SN+CC and SN+CC+H0LiCOW data, respectively. Note that without BAO data these constraints are compatible with each other practically across the whole $z$ range under consideration, where the addition of the H0LiCOW sample, significantly improves the reconstruction compared to SN+CC. It is also interesting to observe the behavior for $z > 1.5$, where we see that $Om < 0.31$. Combining BAO data with SN+CC+H0LiCOW, we observe a significant improvement in the reconstruction for the whole $z$ range considered in the analysis. Predictions for $z > 2$ disagree at $\sim$1.5$\sigma$ CL when the GP mean  is compared between SN+CC+H0LiCOW and SN+CC+BAO+H0LiCOW.

On the other hand, the BAO measurements require a calibration of the sound horizon, either through BBN or the CMB. In all our analyses, we have used BAO data with the assumption $r_d/r_{d, fid} = 1$, where $r_{d,fid}$ is the fiducial input value. In order to quantify how much the $r_d$ value can influence the GP reconstruction, we have analyzed $Om(z)$ with different $r_d$ input values. We have used $r_d$ values obtained from Planck-CMB data \cite{Planck2018} and eBOSS Collaboration \cite{SDSS_final}. Figure \ref{Om_results_appendix_2} shows $O_m(z)$, reconstructed using $r_d = 149.30$ Mpc (eBOSS estimation) and $r_d = 147.09$ Mpc (Planck-CMB estimation). In short, we conclude that appropriate and different $r_d$ input values do not change the results significantly in all the analyses carried out in this work. Any input value of $r_d \in [135, 155]$ Mpc does not have statistical divergence compared to assumption $r_d/r_{d, fid} = 1$. That is, all analyses are consistent with each other at $<$1$\sigma$. Therefore, the GP analyses here are not sensitive to $r_d$. That is why, we have presented the results in the main text assuming $r_d/r_{d, fid} = 1$.

\begin{acknowledgments}
\noindent 
The authors thank to Sunny Vagnozzi, Valerio Marra and  Chris Clarkson for a critical reading of the manuscript and useful comments. S.K. gratefully acknowledges the support from SERB-DST project No. EMR/2016/000258. R.C.N. would like to thank the agency FAPESP for financial support under the project No. 2018/18036-5.  
\end{acknowledgments}

\end{document}